\documentclass{article}

\PassOptionsToPackage{numbers, compress}{natbib}
\usepackage[final]{nips_2016}

\usepackage[utf8]{inputenc} % allow utf-8 input
\usepackage[T1]{fontenc}    % use 8-bit T1 fonts
\usepackage{hyperref}       % hyperlinks
\usepackage{url}            % simple URL typesetting
\usepackage{booktabs}       % professional-quality tables
\usepackage{amsfonts}       % blackboard math symbols
\usepackage{nicefrac}       % compact symbols for 1/2, etc.
\usepackage{microtype}      % microtypography

\usepackage{listings}
\usepackage{color}
\usepackage{subcaption}
\usepackage{booktabs}
\usepackage{url}
\usepackage{amsmath}
\usepackage{graphicx} %,subfigure}
\usepackage[flushleft]{threeparttable}
\usepackage{tabularx}
\usepackage{hhline}
\usepackage{mdwlist}
\usepackage{multirow}
\usepackage{rotating}

\usepackage{algorithm}
\usepackage{algorithmic}

\title{Parallelizing Word2Vec in Multi-Core and Many-Core Architectures}

\author{
  Shihao Ji, Nadathur Satish, Sheng Li, Pradeep Dubey \\
  Parallel Computing Lab, Intel Labs, USA \\
  \texttt{\{shihao.ji, nadathur.rajagopalan.satish, sheng.r.li, pradeep.dubey\}@intel.com} \\
}

\input{Definitions}

\begin{document}

\maketitle

\begin{abstract}
Word2vec is a widely used algorithm for extracting low-dimensional vector representations of words. State-of-the-art algorithms including those by Mikolov et al.~\cite{MikCheCorDea13,MikSutChe13} have been parallelized for multi-core CPU architectures, but are based on vector-vector operations with ``Hogwild" updates that are memory-bandwidth intensive and do not efficiently use computational resources. In this paper, we propose ``HogBatch" by improving reuse of various data structures in the algorithm through the use of minibatching and negative sample sharing, hence allowing us to express the problem using matrix multiply operations. We also explore different techniques to distribute \wv computation across nodes in a compute cluster, and demonstrate good strong scalability up to 32 nodes. The new algorithm is particularly suitable for modern multi-core/many-core architectures, especially Intel's latest Knights Landing processors, and allows us to scale up the computation near linearly across cores and nodes, and process hundreds of millions of words per second, which is the fastest \wv implementation to the best of our knowledge. 
\end{abstract}

\section{From Hogwild to HogBatch}
\label{sec:algorithm}
%In order to solve the optimization problem described in the previous
%section, Stochastic Gradient Descent (SGD) is commonly used.
%SGD is an iterative algorithm; at each
%iteration, a single $(w_I, w_O)$ pair is picked, where $w_I$ is an input
%context word and $w_O$ is a target word or a negative sample. The gradient of
%the objective function is then calculated w.r.t. the word
%vectors of $w_I$ and $w_O$; and a small change/update is made to these
%vectors. One of the problems of SGD is that it is inherently challenging to parallelize, i.e., SGD only
%updates the word vectors of a pair of words at a time, and parallel model
%updates on multiple threads can result in conflicts if the threads try
%to update the vectors of same words.

\begin{figure*}[htb]
\centering
\includegraphics[width=5.0in]{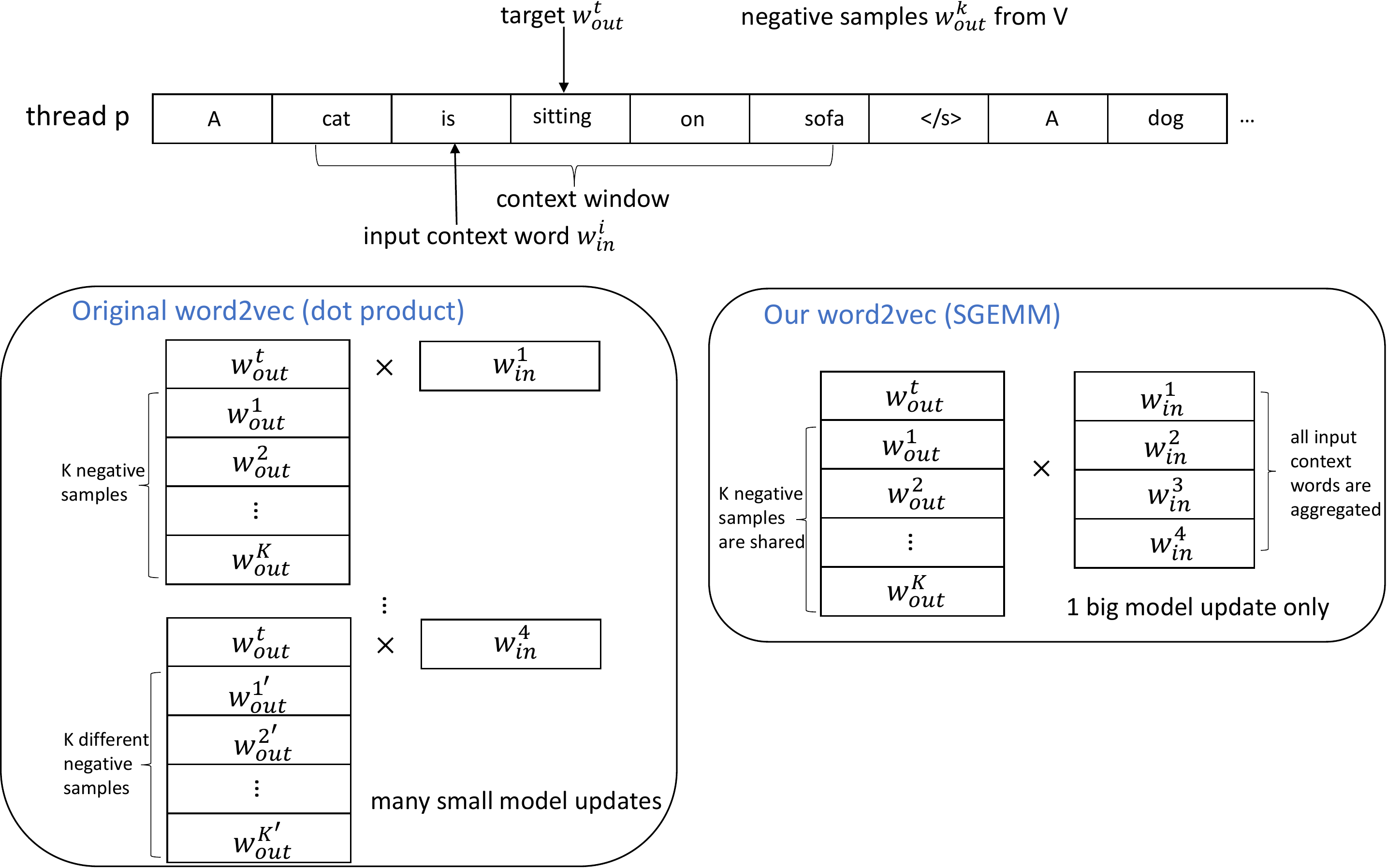} \vspace{-0.1cm}
\caption{The parallelization schemes of the original word2vec (left) and our optimization (right).}
\label{fig:schemes}
\vspace{-0.2cm}
\end{figure*}

We refer the reader to~\cite{MikCheCorDea13,MikSutChe13} for an introduction to \wv and its optimization problem. The original implementation of \wv by Mikolov \emph{et
al.}~\footnote{\url{https://code.google.com/archive/p/word2vec/}} uses
Hogwild~\cite{NiuRecRe11} to parallelize SGD. 
Hogwild is a parallel SGD algorithm that seeks to ignore
conflicts between model updates on different threads and allows updates to proceed even in the presence of conflicts.
The psuedocode of \wv Hogwild SGD is shown in
Algorithm~\ref{alg:hogwild}. The algorithm takes in a matrix $M_{in}^{V\times D}$ that
contains the word representations for each input word, and a matrix $M_{out}^{V\times D}$
for the word representations of each output word.
Each word is represented as an array of $D$ floating point numbers,
corresponding to one row of the two matrices.
These matrices are updated during the training. We take in a target word, and a set of $N$ input context words around the target as depicted in the top of Figure~\ref{fig:schemes}.
The algorithm iterates over the $N$ input words in Lines 2-3. In the loop at Line 6, we pick either the positive
example (the target word in Line 8) or a negative example at random (Line
10). Lines 13-15 compute the gradient of the objective function with
respect to the choice of input word and positive/negative example. Lines
17--20 perform the update to the entries $M_{out}[\text{pos/neg example}]$
and $M_{in}[\text{input context}]$. The psuedocode only shows a single thread; in Hogwild, the loop in Line 2 is parallelized over threads without any additional change in the code. 

%\begin{table}
%\captionof{algorithm}{Hogwild SGD implementation of \wv in one thread.}
%\label{alg:hogwild}
%\begin{minipage}{0.01\linewidth}\hspace{0.01pt}\end{minipage}
%\begin{minipage}{0.95\linewidth}
%\begin{tabular}{l}
%%\toprule
%\begin{lstlisting}
%@Given model parameter $\Omega=\{M_{in},M_{out}\}$, learning rate $\alpha$, 1 target word $w_{out}^t$, and N input words  \{$w_{in}^0$, $w_{in}^1$, $\cdots$, $w_{in}^{N-1}$\}@
%for (i = 0; i < N; i++) {
%  input_word = @$w_{in}^i$@;
%  for (j = 0; j < D; j++) temp[j] = 0;
%  // negative sampling
%  for (k = 0; k < negative + 1; k++) {
%    if (k = 0) {
%      target_word = @$w_{out}^t$@; @label@ = 1;
%    } else {
%      target_word = sample one word from V; @label@ = 0;
%    }
%    inn = 0;
%    for (j = 0; j < D; j++) inn += @$M_{in}$@[input_word][j] * @$M_{out}$@[target_word][j];
%    err = @label@ - @$\sigma$@(inn);
%    for (j = 0; j < D; j++) temp[j] += err * @$M_{out}$@[target_word][j];
%    // update output matrix
%    for (j = 0; j < D; j++) @$M_{out}$@[target_word][j] += @$\alpha$@ * err * @$M_{in}$@[input_word][j];
%  }
%  // update input matrix
%  for (j = 0; j < D; j++) @$M_{in}$@[input_word][j] += @$\alpha$@ * temp[j] ;
%}
%\end{lstlisting}\\
%%\bottomrule
%\end{tabular}
%\end{minipage}
%\end{table}

\begin{algorithm}
  \begin{algorithmic}[1]
    \STATE {Given model parameter $\Omega=\{M_{in},M_{out}\}$, learning rate $\alpha$, 1 target word $w_{out}^t$, and N input words  \{$w_{in}^0$, $w_{in}^1$, $\cdots$, $w_{in}^{N-1}$\}}
    \STATE {for (i = 0; i < N; i++) \{}
    \STATE {\hspace{0.5cm}input\_word = $w_{in}^i$;}
    \STATE {\hspace{0.5cm}for (j = 0; j < D; j++) temp[j] = 0;}
    \STATE {\hspace{0.5cm}// negative sampling}
    \STATE {\hspace{0.5cm}for (k = 0; k < negative + 1; k++) \{}
    \STATE {\hspace{1.0cm}if (k = 0) \{}
    \STATE {\hspace{1.5cm}target\_word = $w_{out}^t$; label = 1;}
    \STATE {\hspace{1.0cm}\} else \{}
    \STATE {\hspace{1.5cm}target\_word = sample one word from V; label = 0;}
    \STATE {\hspace{1.0cm}\}}
	\STATE {\hspace{1.0cm}inn = 0;}
    \STATE {\hspace{1.0cm}for (j = 0; j < D; j++) inn += $M_{in}$[input\_word][j] * $M_{out}$[target\_word][j];}
    \STATE {\hspace{1.0cm}err = label - $\sigma$(inn);}
    \STATE {\hspace{1.0cm}for (j = 0; j < D; j++) temp[j] += err * $M_{out}$[target\_word][j];}
    \STATE {\hspace{1.0cm}// update output matrix}
    \STATE {\hspace{1.0cm}for (j = 0; j < D; j++) $M_{out}$[target\_word][j] += $\alpha$ * err * $M_{in}$[input\_word][j];}
    \STATE {\hspace{0.5cm}\}}
    \STATE {\hspace{0.5cm}// update input matrix}
    \STATE {\hspace{0.5cm}for (j = 0; j < D; j++) $M_{in}$[input\_word][j] += $\alpha$ * temp[j];}
    \STATE{\}}
  \end{algorithmic}
  \caption{\wv Hogwild SGD in one thread.}
  \label{alg:hogwild}
\end{algorithm} %\vspace{-0.2cm}

Algorithm~\ref{alg:hogwild} reads and updates
entries corresponding to the input context and positive/negative words at
each iteration of the loop at Line 6. This means that there is a
potential dependence between successive iterations - they may happen to touch the same word representations, and each iteration must potentially wait for the update from the previous iteration to complete. Hogwild ignores
such dependencies and proceeds with updates regardless of conflicts.
In theory, this can reduce the rate of convergence of the algorithm as
compared to a sequential run. However, the Hogwild approach has been
shown to work well in case the updates across
threads are unlikely to be to the same word; and indeed for large
vocabulary sizes, conflicts are relatively rare and convergence is not
typically affected.

%\subsection{Advantages and Drawbacks of Algorithm~\ref{alg:hogwild}}
%Algorithm~\ref{alg:hogwild} has a few main advantages: threads do not
%need to synchronize between updates and can hence proceed
%independently with minimal instruction overheads. Further, the
%computation of the gradient is based off the current state of the
%model visible to the thread at that time. Since all threads update the
%same shared model, the values read are only as stale as the
%communication latency between threads, and in practice this does not
%cause much convergence problems for \wv.

\vspace{-0.3cm}

\subsection{Shared Memory Parallelization: HogBatch}

%We first discuss how we can exploit the available locality in
%Algorithm~\ref{alg:hogwild}. This can be done even on a single compute
%thread. We then describe the impact of this step on parallelization
%and inter-thread communication.

However, the original \wv algorithm suffers from two main drawbacks that
significantly affect runtimes. First, since
multiple threads can update the same cache line containing a specific
model entry, there can be significant ping-ponging of cache lines
across cores. This leads to high access latency and significant drop
in scalability. Second and perhaps even more importantly, there is
a significant amount of locality in the model updates that is not
exploited in the Hogwild algorithm. As an example, we can easily see
that the same target word $w_{out}^t$ is used in the model updates for several
input words. By performing a single update at a time, this locality
information is lost, and the
algorithm performs a series of dot-products that are level-1 BLAS operations~\cite{BlaDemDon02} and limited
by memory bandwidth. It is indeed, as we show next, possible to batch
these operations into a level-3 BLAS call~\cite{BlaDemDon02} which can more efficiently utilize
the compute capabilities and the instruction sets of modern multi-core
and many-core architectures.

We exploit locality in two steps. As a motivation, consider
Figure~\ref{fig:schemes}. The figure to the left shows the
parallelization scheme of the original \wv. Note that we compute
dot products of the word vectors for a given input word $w_{in}^{i}$
with both the target word $w_{out}^{t}$ as well as a set of $K$ negative samples
$\{w_{out}^{1}, \cdots, w_{out}^{K}\}$. Rather than doing these one at a
time, it is rather simple to batch these dot products into a matrix
vector multiply, a level-2 BLAS operation~\cite{BlaDemDon02}, as shown in the left side of Figure~\ref{fig:schemes}.
However, this alone does not buy significant performance improvement.
Indeed, most likely the shared input word vector may come from cache. In order to convert this to a level-3 BLAS operation, we also need to batch
the input context words. Doing this is non-trivial since the negative
samples for each input word could be different in the original
\wv implementation. We hence propose ``negative sample sharing'' as a strategy,
where we share negative samples across a small batch of input words.
Doing so allows us to convert the original dot-product based
multiply into a matrix-matrix multiply call (GEMM) as shown on the right side of
Figure~\ref{fig:schemes}. At the end of the GEMM, the model
updates for all the word vectors of all input words and target/sample
words that are computed need to be written back. Performing
matrix-matrix
multiplies (GEMMs) rather than dot-products allows us to leverage
all the compute capabilities of modern architectures including vector units and
instruction set features such as multiply-add instructions in the
Intel %{Intel$^{\mbox{\tiny\textregistered}}$} 
AVX2 instruction set.
It also allows us to leverage heavily optimized linear algebra
libraries. 
%Note that typical matrix dimensions are not very large. For
%instance, the number of negative samples is only 5--20, and the batch
%size for the input batches are limited to about 10--20 for convergence
%reasons. Nevertheless, we find that we get considerable speedups even
%with this level of reuse over the original \wv.

For multi-threading across the GEMM calls, we follow the same ``Hogwild"-style
philosophy - each thread performs its own GEMM call independently to other threads, and we
allow for threads to potentially conflict when updating the models at
the end of the GEMM operation. We therefore call our new parallelization scheme ``HogBatch".

%\subsection{Consequence of the new parallelization scheme}
%%\subsubsection{Multithreading}
While the original \wv performs model updates after each dot product, our HogBatch
scheme performs a number of dot products as a GEMM call before
performing model updates. It is important to note that this locality optimization has a secondary but important benefit - we cut down on the total number of updates to the model.
This happens since the GEMM operation performs a reduction
(in registers/local cache) to an update to a single entry in the
output matrix; while in the original \wv scheme such updates to the same
entry (same input word representation, for instance) happen at
distinct periods of time with potential ping-pong traffic happening
in between. As we will see in Sec.~\ref{sec:experiments} when we present results, this leads to
a much better scaling of HogBatch than the original \wv.

\subsection{Distributed Memory Parallelization}
%Scalability on multi-node distributed system is as important as, if not more important than, that on single node system. This is because typical large scale machine learning applications are compute intensive and require days, weeks even months of training time. In the case of \wv, even with the techniques we proposed above, it still takes tens of hours or even days to train on some of the largest data sets in the industry, such as the 100 billion word news articles from Google. Thus, scaling out \wv on multi-node distributed system is critical in practice.

To scale out \wv, we also explore different techniques
to distribute its computation across nodes in a compute cluster. Essentially, we employ data parallelism for distributed computation. Due to limited space, we skip the details here and will report it in a full paper.

\section{Experiments}
\label{sec:experiments}

%We optimize \wv with the techniques discussed above both in single node shared memory system and in multi-node distributed system. We report the \emph{system-performance} measured as throughput, i.e., million words/sec, and the \emph{predictive-performance} measured as accuracy on standard word similarity and word analogy test sets. The performances of our optimization are compared with the original \wv on CPUs, and with the state-of-the-art results reported in literature on Nvidia GPUs.

%With the techniques and optimizations discussed above, our optimized \wv delivers the highest \emph{system-performance}, measured as throughput, i.e., million words/sec, reported to date on both single node shared memory systems and multi-node distributed clusters, while maintaining \emph{predictive-performance} measured as accuracy. This section provides detailed analysis of our algorithm on shared memory and distributed memory systems.

%In particular, compared with all state-of-the art \wv designs, our distributed \wv achieves a record-setting throughput of 110 million words/sec on a cluster of 32 nodes, while maintaining accuracy.

%\subsection{Experimental Setup}
We compare the performances of three different implementations of \wv: (1) the original implementation from Google that is based on Hogwild SGD on shared memory systems (\url{https://code.google.com/archive/p/word2vec/}), (2) BIDMach (\url{https://github.com/BIDData/BIDMach}) which achieves the best known performance of \wv on Nvidia GPUs, and (3) our optimized implementation on Intel architectures, including (1) 36-core Intel Xeon E5-2697 v4 Broadwell (BDW) CPUs, and (2) the latest Intel Xeon Phi 68-core Knights Landing (KNL) processors. We train the algorithm on the one billion word benchmark~\cite{CheMikSch14} with the same parameter settings of BIDMatch (dim=300, negative samples=5, window=5, sample=1e-4, vocabulary of 1,115,011 words). We evaluate the model accuracy on the standard \emph{word similarity} benchmark WS-353~\cite{FinGabMat02} and Google \emph{word analogy} benchmark~\cite{MikCheCorDea13}. Since all the implementations achieve similar accuracy and due to lack of space, in the following we only report their performances in term of throughput, measured as million words/sec. More details of the experimental comparison will be reported in a full paper. Our implementation and scripts are open sourced at \url{https://github.com/IntelLabs/pWord2Vec}.

\begin{figure}[htb]
\centering
\includegraphics[width=2.4in]{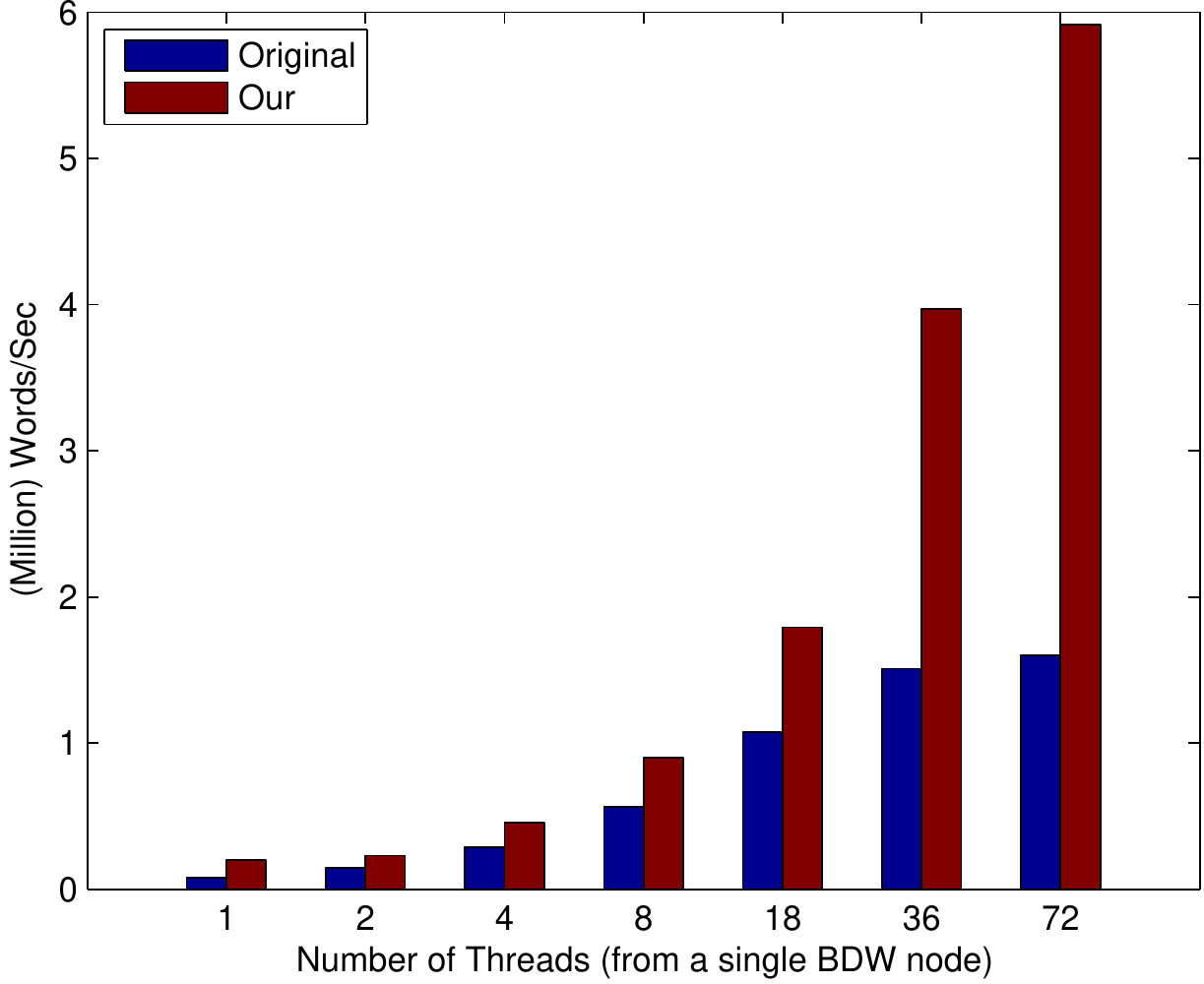} \hspace{1.0cm}
\includegraphics[width=2.5in]{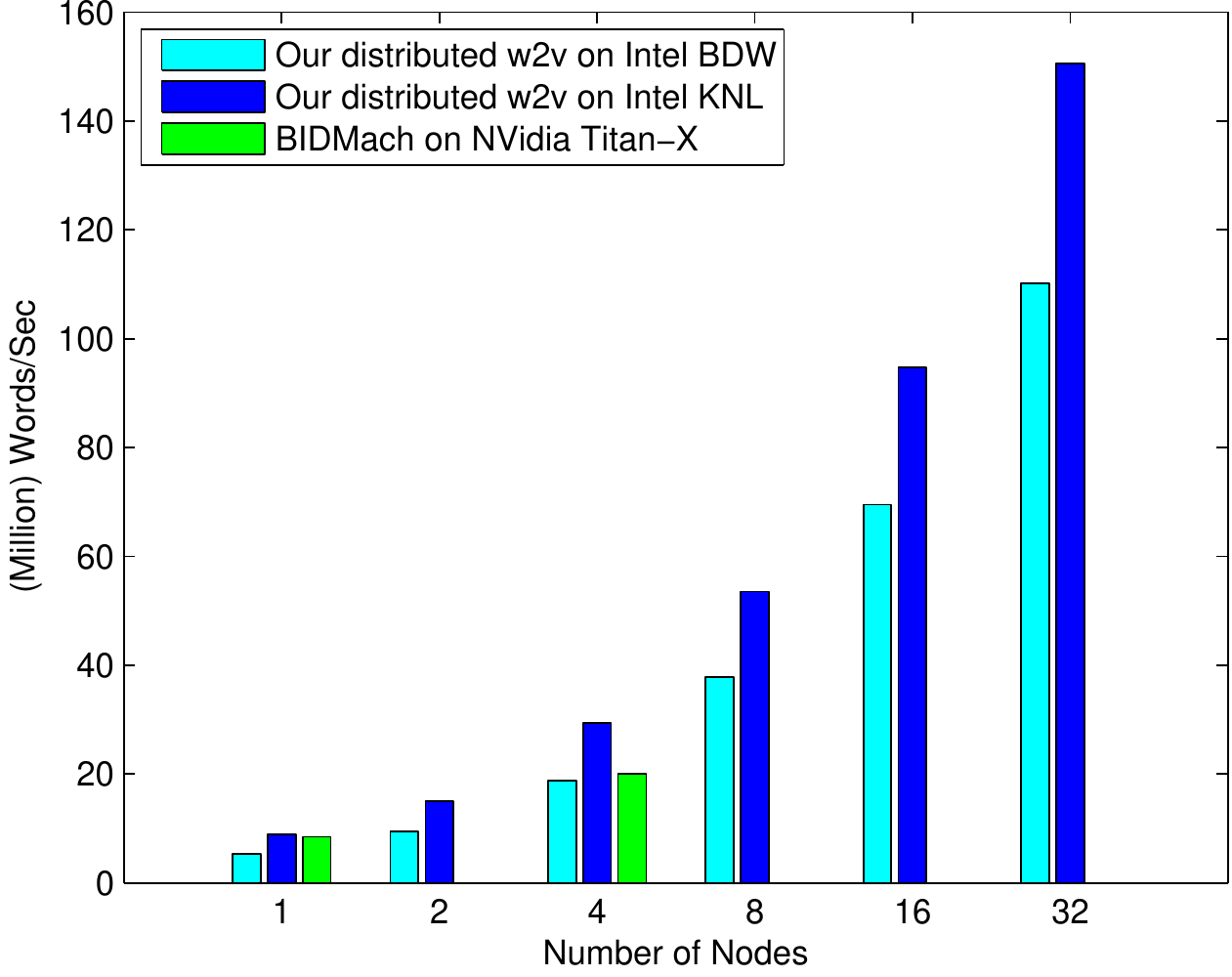} \vspace{-0.2cm}
\caption{(a) Scalabilities of the original \wv and our optimization on all threads of an Intel Broadwell CPU; (b) Scalabilities of our distributed \wv on multiple Intel Broadwell and Knights Landing nodes, and BIDMach on $N=1,4$ NVidia Titan-X nodes as reported in \cite{CanZhaChe15}.}
\label{fig:scaling}
\vspace{-0.1cm}
\end{figure}

Figure~\ref{fig:scaling} shows the throughputs measured as million words/sec of our algorithm and the original \wv, scaling across cores and nodes of Intel BDW and KNL processors. When scaling to multiple threads (Figure~\ref{fig:scaling}(a)), our algorithm achieves near linear speedup until 36 threads. In contract, the original \wv scales linearly only until 8 threads and slows down significantly after that. In the end, the original \wv delivers about 1.6 million words/sec, while our code delivers 5.8 million words/sec or a 3.6X speedup over the original \wv. The superior performance highlights the effectiveness of our optimization in reducing unnecessary inter-thread communications and utilizing computation resource of modern multi-core architecture. When scaling across multiple nodes (Figure~\ref{fig:scaling}(b)), our distributed \wv achieves near linear scaling until 16 BDW nodes or 8 KNL nodes while maintaining a similar accuracy to that of the original \wv. As the number of nodes increases, to maintain a comparable accuracy, we need to increase the model synchronization frequency to mitigate the loss of convergence rate. However, this takes a toll on the scalability and leads to a sub-linear scaling at 32 BDW nodes or 16 KNL nodes. Despite of this, our distributed \wv delivers over 100 million words/sec with a small 1\% accuracy loss. To the best of our knowledge, this is the best performance reported so far on this benchmark. Finally, Table~\ref{tab:throughputs} summarizes the best performances of the state-of-the-art implementations on different architectures, demonstrating superior performance of our algorithm.

\begin{table}[htb]   \vspace{-0.2cm}
  \caption {Performance comparison of the state-of-the-art implementations of \wv on different architectures.} \vspace{-0.2cm}
  \label{tab:throughputs}
  \begin{center}
	\begin{threeparttable}
    \begin{tabular}{|l|l|c|}
      \hline Processor      & Code  & Words/Sec \\
%      \hhline{|=|=|=|} Intel HSW (Xeon E5-2680 v3) & Original & 1.5M \\
%      \hline Intel HSW (Xeon E5-2680 v3) & BIDMach & 2.4M \\
%      \hline Intel HSW (Xeon E5-2680 v3) & Our & \textbf{4.2M} \\
      \hhline{|=|=|=|} Intel BDW (Xeon E5-2697 v4) & Original & 1.6M \\
      \hline Intel BDW (Xeon E5-2697 v4) & BIDMach & 2.5M \\
      \hline Nvdia K40 & BIDMach & \hspace{0.05in}4.2M$^1$ \\
      \hline Intel BDW (Xeon E5-2697 v4) & Our & \textbf{5.8M} \\
      \hline Nvdia GeForce Titan-X & BIDMach & \hspace{0.05in}8.5M$^1$ \\
      \hline Intel KNL & Our & \textbf{8.9M} \\
      \hhline{|=|=|=|} Nvdia GeForce Titan-X (4 nodes) & BIDMach & \hspace{0.05in}20M$^1$ \\
      \hline Intel KNL (4 nodes) & Our & \textbf{29.4M} \\      
      \hline
    \end{tabular}%\vspace{-0.3cm}
    \begin{tablenotes}
  		\scriptsize\item $^1$Data from \cite{CanZhaChe15}.
	\end{tablenotes}
	\end{threeparttable}
  \end{center}\vspace{-0.6cm}
\end{table}

\section{Conclusion}
\label{sec:conclusion}

A high performance \wv algorithm ``HogBatch" is proposed for shared memory and distributed memory systems. The algorithm is particularly suitable for modern multi-core/many-core architectures, especially Intel's KNL, on which we deliver the best known performance reported so far. Our implementation is publicly available for general usage. 

%It combines the idea of Hogwild, minibatching and shared negative sampling to convert the level-1 BLAS vector-vector operations to the level-3 BLAS matrix multiply operations. As a result, the proposed algorithm is more hardware-friendly and can efficiently leverage the vector units and multiply-add instruction of modern multi-core and many-core architectures. We also explore different techniques, such as sub-model synchronization and learning rate scheduling, to parallelize the \wv computation across multiple computing nodes. These techniques dramatically reduce network communication and keep the model synchronized effectively when number of nodes increases. 
%%We demonstrate the throughput and predictive accuracy of our algorithm comparing to the state-of-the-arts implementations, such the original \wv and BIDMach, on both single node shared memory systems and multi-node distributed systems. 
%We achieve near linear scalability across cores and nodes, and process hundreds of mullions of words per second, the best performance reported so far on the one billion word benchmark.

%As for future work, our plans include asynchronous model update similar to parameter sever~\cite{LiAndSmo14}, more efficient sub-model synchronization strategy as well as improving the rate of convergence of the distributed \wv implementation.

%\section*{Acknowledgment}

%\bibliographystyle{unsrt}{abbrvnat}
\bibliographystyle{abbrvnat}
\bibliography{ref}

\end{document}